# A spinal circuit for collective coordination

L.D. Picton[1*#], D. Madrid[1*], A. Pazzaglia[2*], Y. Wang[3], M. Bertuzzi[1], A. Ferrario[2], A. Anastasiadis[2,4] J. Arreguit[2], P. Fontanel[1], C.-X. Huang[3], K. Mulleners[4], J. Song[1,3#], A. Ijspeert[2#], A. El Manira[1#]

[1]Department of Neuroscience, Karolinska Institutet, Stockholm, Sweden

[2]Biorobotics Laboratory, École Polytechnique Fédérale de Lausanne (EPFL), Lausanne, Switzerland

[3]Shanghai Key Laboratory of Anesthesiology and Brain Functional Modulation, Clinical Research Center for Anesthesiology and Perioperative Medicine, Translational Research Institute of Brain and Brain-Like Intelligence, Shanghai Fourth People's Hospital, School of Medicine, Tongji University, Shanghai 200434, China

[4]Unsteady Flow Diagnostics Laboratory, Institute of Mechanical Engineering, École Polytechnique Fédérale de Lausanne (EPFL), Lausanne, Switzerland

[*]Equal contributions

[#]corresponding authors

**Abstract**

The coordinated movement of animal groups is one of the most widespread social behaviors, which are generally attributed to high-order cognitive processing in the brain. Yet, collective coordination can seemingly emerge from rapid, local interactions between individuals, suggesting the existence of decentralized mechanisms of online coordination that remain to be identified. Here, we show that a low-order spinal sensorimotor circuit is required for real-time social coordination during schooling in zebrafish. Central to this circuit are intraspinal proprioceptive neurons that detect local body bending and deliver direct, curvature-based inhibition to precisely time the locomotor network. Combining electrophysiology, calcium imaging, optogenetics, and behavioral analysis, we show that this circuit encodes both self-generated (egocentric) and neighbor-induced (allocentric) body bending signals, enabling fish to match the phase of their swimming to the wakes of their neighbors (vortex phase matching). In a neuromechanical model and physical robot, this single feedback loop is sufficient to generate vortex phase matching and to lower the energetic cost of swimming. Disrupting this circuit uncouples neighboring fish and abolishes schooling behavior. These results show that a spinal circuit dynamically synchronizes individuals through simple, local interactions, revealing how low-order mechanisms can drive the emergence of coordinated group behavior.

**Introduction**

*No man is an island* [1]; the behavior of an individual is deeply embedded within collective dynamics [2-7]. Across the animal kingdom, individuals moving together produce ordered collective patterns: starlings in flocks, geese moving in V-formations, and spontaneous flow in dense human crowds [8,9]. The most widespread example is fish schooling. Across the planet, trillions of fish spend much of their lives in dynamic, swirling formations in which the school moves as a unified entity, enhancing survival, foraging, navigation and energy saving [10].

Social behaviors have been widely studied in the context of complex high-order integration and processing in the brain [2,3,5,6,11-13]. Yet, effective group coordination requires mechanisms for rapid, decentralized responses to cues generated by nearby individuals [14,15]. A low-order conceptual model suggests that animals can achieve this by following simple local rules, leading to the emergence of fast and adaptive social coordination without the need for higher-order computation [16,17]. Such coordination reflects the general principle of synchronization, where simple local interactions among coupled oscillators can give rise to coherent collective patterns [18]. Within a school, each fish is coupled mechanically to its neighbors through the wakes they leave behind, just as birds in formation fly in the air currents of those ahead. What has been missing is the biological identity of the sensor that detects a neighbor's hydrodynamic signal, and the circuit that converts it into a synchronization of motor behavior.

Here we reveal a localized, low-order spinal circuit mechanism that underpins real-time collective coordination during zebrafish schooling. In this circuit, intraspinal proprioceptors detect local body bending and exert online control over the excitatory core of the locomotor circuit, enabling individuals to sense and respond dynamically to the vortex trails of neighboring fish. We show that this sensorimotor circuit underlies real-time detection and integration of hydrodynamic cues, driving the phase coupling of swimming between neighboring individuals and the emergence of coordinated collective behavior.

## Materials and Methods

*Animals*

Experiments were conducted using wild-type and genetically modified zebrafish (*Danio rerio*). Animals were raised and maintained in a core facility at the Karolinska Institute, following established protocols. In this study, both juvenile (4-6 weeks old) and adult zebrafish (8-10 weeks old) of either sex were used. All experimental procedures were approved by the local Animal Research Ethical Committee, Stockholm and were performed in accordance with EU guidelines.

*Zebrafish preparation for electrophysiological experiments*

Adult zebrafish (8-10 weeks old) were dissected following the *ex-vivo* brainstem-spinal cord protocol previously published [19-21]. The *ex-vivo* preparation was then transferred to the recording chamber where it was embedded on 2% low-melting agarose (Sigma-Aldrich) to ensure stability. The tail of the fish was released from the agarose to allow self-generated movement or imposed manipulation. Additionally, 5-8 segments of the dissected spinal cord region were exposed to enable access with the intracellular electrode. The preparation was kept at room temperature (20-22°C) and continuously perfused with oxygenated extracellular solution composed of (in mM): 134 NaCl, 2.9 KCl, 2.1 $CaCl_2$, 1.2 $MgCl_2$, 10 HEPES, and 10 glucose with pH 7.8 adjusted using NaOH and an osmolarity of approximately 290 mOsm.

*Patch-clamp electrophysiology*

Whole-cell patch-clamp electrophysiological experiments were conducted using glass electrodes made from borosilicate glass (Hilgenberg) and pulled with a micropipette puller (P-1000, Sutter Instruments). The electrodes were filled with an intracellular solution containing (in mM): 120 K-gluconate, 5 KCl, 10 HEPES, 4 $Mg_2ATP$, 0.3 $Na_4GTP$, and 10 Na-phosphocreatine, with the pH adjusted to 7.4 using KOH and an osmolarity of 275 mOsm. Neurons were visualized with a fluorescence microscope (Axioskop FS Plus, Zeiss) equipped with IR-differential interference contrast (DIC) optics and a CCD camera (Hamamatsu). An extracellular recording electrode (EMG electrode) was positioned on the muscle. Extracellular signals were amplified with a differential AC amplifier (A-M Systems) and filtered using low and high cut-off frequencies of 300 Hz and 1 kHz, respectively. Neuronal electrophysiological data were analyzed using Clampfit (Molecular Devices) and a MATLAB custom script (MATLAB 2021b, MathWorks).

*Mechanical tail manipulation experiments*

To control movement of the tail of the fish during optogenetically induced swimming in head-fixed preparations without skin, the tail was connected to a servomotor (Dynamixel MX-12W, Robotis Inc.) controlled by an Arduino Uno (Arduino LLC) using the *DynamixelShield.h* library. For experiments testing the effects of egocentric signaling, the tail was either free to move naturalistically during swimming, or movement was blocked during swimming. During static bending experiments, lateral displacements were applied either to the left or to the right during swimming for approximately 4 seconds, before returning the tail to a straight position. In *ex vivo* pacing experiments, rhythmic tail bending sequences with amplitude matching those observed during normal swimming were applied at fixed frequencies (2, 4, 6, 8, and 10 Hz) were programmed using the Arduino IDE (Arduino LLC). For *ex vivo* schooling experiments, a consecutive sequence of tail kinematic values was extracted from the behavior of a leader fish during a close-pair interaction in our behavioral data.

*Calcium imaging*

For calcium imaging experiments, the following zebrafish lines were used*:* Tg(*chx10*:gal4) x Tg(uas:gcamp6s) or Tg(*chx10*:gal4) x Tg(uas:gcamp6s) x Tg(uas:chr2-yfp) to visualize V2a interneuron activity and Tg(*glyt2*:gal4) x Tg(uas:gcamp6s) [22,23] to visualize intraspinal proprioceptor activity. All calcium imaging experiments were conducted using a laser confocal microscope (Zeiss LSM 980-Airy) equipped with a 25x water immersion objective.

*Behavior*

Schooling behavior experiments were performed using a custom-built swim tunnel [24]. Groups of either 7-11 fish, or pairs of fish (6-8 weeks old), were placed in the swim tunnel, and were recorded using a high-speed camera (GoPro HERO9) at 240 fps. The flow was set to 10-12 cm/s, which corresponds to a medium range speed tolerated by the fish (~60% of maximum swim speed). For schooling behavior experiments in static water animals were placed in a dish (diameter 19 cm) containing fish water positioned on a Plexiglass platform, illuminated from below by a light-emitting diode lightbox and imaged from above with a high-speed camera. Groups of 8 zebrafish were placed in the dish and were recorded using a high-speed camera at 187 fps after being allowed to habituate to the arena for at least 15 minutes.

*Behavioral data analysis*

Behavioral videos were analyzed using DeepLabCut (v.3.0.0rc8) and custom MATLAB scripts (MATLAB 2021b, MathWorks). For experiments involving groups of fish in the flow tunnel, the position of each animal was extracted as its centroid. For each frame, centroid coordinates were ordered by their angular orientation relative to the group's geometric center to form a closed polygon with no self-intersections. Polygons were generated for each experimental condition and aligned to the group center of mass to facilitate comparison. For experiments with pairs of fish in the flow tunnel, the kinematics and position of each fish were extracted from a labeled skeleton (3 labels: head, center and tail). The stabilized centroid coordinates were used to determine the position of each fish and identify the leader, defined as the fish within the pair with the smallest x-coordinate relative to the origin of the current (on the left side of the tank). The angle ($\Psi$) of the tail was computed by summing the angles of the vectors connecting the skeleton. Once the angles were extracted for each frame, the cycle phase of the leader ($\Phi L$) and follower ($\Phi F$) was calculated by mapping the frame numbers between consecutive peaks of the trace (where each peak represents the maximum body bend in the same direction) to an interval of $[0, 2\pi]$.

*Computational and mechanical model*

To simulate the rhythmic activity observed in the zebrafish spinal cord during locomotion, we developed a spiking neural network based on adaptive exponential integrate-and-fire neuron models [25]. The neural model consists of 32 segments [26], each containing pools of excitatory V2a interneurons and inhibitory V0d interneurons, which together form the core of the locomotor CPG [27-29]. The V2a interneurons provide rhythm generation via recurrent excitatory connections, while V0d interneurons mediate left-right alternation through inhibitory commissural projections. Additionally, the network comprises motor neurons, reticulospinal neurons, and intraspinal proprioceptive neurons [27,30,31]. The reticulospinal neurons deliver descending drive uniformly across the CPG segments, activating locomotor rhythms, while the proprioceptive neurons relay sensory feedback related to axial body stretch [31].

The neural network interfaces with a biomechanical model of the zebrafish body comprising 16 rigid links connected by 15 muscle-controlled joints, with motor neuron activity converted into joint torques through an Ekeberg muscle model. The muscle model includes active contraction, passive stiffness, active stiffness, and damping (Supplementary Material). The body interacts with water through an inertial drag-based hydrodynamic model, with parameters chosen to replicate zebrafish swimming performance [32,33]. Neural simulations were performed using the Brian2 library [34,35], while the biomechanical simulations employed the MuJoCo physics engine [36] integrated via the FARMS framework [37]. Fluid dynamics were modeled based on WaterLily.jl [38].

*Robotic fish*

The robot, 1-guilla, is inspired by the morphology of undulatory swimmers [39] . The body of the robot weighs 1.30 kg, measures L = 0.85 m in length with a cross-section of 3.5 cm × 4.35 cm (w × h), and consists of a head, eight actuated body segments, and a tail. The tail is passive and flexible, and it is magnetically attached to the last segment. The segments of the body are linked by custom 3D-printed rigid parts, with each segment actuated by a Dynamixel XM430-W210-R servo motor. An external Raspberry Pi 5 board, interfaced with a Dynamixel U2D2, controls the motors in current control mode. Power was supplied to the motors and the board by a DC power source (12V, max 3A). The controller of the robot was implemented in C++ to replicate the same spiking neural network running in the neuromechanical simulations. The stretch feedback information provided to the spiking neural network was the angular positions of the motors ($\theta_i$), obtained from the encoder readings. The wake of a leading swimmer was replicated by a flapping airfoil placed upstream of the robot. The motion of the robot and the flapper was captured by a camera mounted below the water channel.

*Generation of piezo2b mutant zebrafish*

A piezo2b^ins17/ins17 zebrafish mutant was generated using CRISPR/Cas9 genome editing, following protocols established by Varshney et al., 2015 and 2016 [40,41], and implemented at the DanioReadout (formerly Genome Engineering Zebrafish, SciLifeLab) Facility in Uppsala, Sweden. Wild-type AB embryos were injected at the one-cell stage with 25–50 pg of synthetic sgRNA and 150–300 pg of Cas9 mRNA per embryo, as described in Gudmundsson et al., 2019 [42].

*Conditional ablation of intraspinal proprioceptors and characterization*

For Cre-dependent ablation of centrally located, piezo2b-expressing proprioceptors, embryos from a Tg(glyt2:loxP–DsRed–loxP–GFP) × Tg(dbx1b:Cre) cross were injected at the one-cell stage with the pTol2–piezo2b:loxP–STOP–loxP–DTA plasmid DNA (20 ng/µl) together with Tol2 transposase mRNA (20 ng/µl). Injections were performed using pulled glass capillaries mounted on a WPI PV830 microinjector, calibrated to deliver ~1 nl per pulse. The piezo2b promoter fragment used to drive the loxP–STOP–loxP–DTA construct spans 5,000 bp immediately upstream of the piezo2b transcription start site (Assembly GRCz11/danRer11; chr2:54,081,433–54,086,433).

*RNAscope in situ hybridization*

Spinal cords were dissected after terminal anaesthesia with 0.1% MS-222 and fixed in 4% paraformaldehyde (PFA) in PBS (0.01 M, pH 7.4) for 24 h at 4 °C. Samples were washed three times for 5 min in PBS before downstream processing. Detection of piezo2b mRNA was performed using the RNAscope Multiplex Fluorescent v2 assay (Advanced Cell Diagnostics) according to the manufacturer’s instructions with minor modifications, following the same workflow described previously for RNAscope in zebrafish spinal cord [31,43].

*Quantification and statistical analysis*

Statistical analyses and data visualization were performed using GraphPad Prism software (https://www.graphpad.com/) or MATLAB. Depending on the comparison, we applied one-way ANOVA with appropriately corrected multiple comparisons, paired or unpaired two-tailed Student's t-tests, Wilcoxon signed-rank test, linear-circular correlation analysis, explained circular variance, or $R^2$-permutation tests (see below), as appropriate. Statistical significance was defined as $p < 0.05$.

## Results

### Dynamics of behavioral coordination during schooling

During schooling, it has been proposed that fish exploit local hydrodynamic perturbations generated by their neighbors to coordinate their swimming movements [44] (Fig. 1A). Each individual receives sensory feedback from their self-generated movements (egocentric signals) while also experiencing body deflections caused by vortex trails from nearby fish (allocentric signals) (Fig. 1B). To investigate whether and how zebrafish school, we analyzed the behavior of groups of adult zebrafish in a swim tunnel (Fig. 1C). They formed groups that consistently maintained tight spatial proximity, forming aggregated constellations (Fig. 1D, green). Another important feature of schooling is the polarization of the group swimming behavior, i.e. the tendency of individuals in a group to swim in the same direction. In static water, zebrafish spontaneously formed and maintained cohesive groups that swam in coordinated and highly polarized schooling constellations, defined by strong alignment of individual swimming directions (Fig. 1, E to G).

To determine whether neighboring fish within a school actively coordinate and synchronize their swimming, we performed detailed pairwise analysis of fish in the swim tunnel and static water conditions. We quantified the front-back distance and relative phase of swim movements of neighboring fish (Fig. 1, H and I), which showed a clear relationship between the phase of the tail angles of neighboring fish (Fig. 1I). Furthermore, in both conditions, the relative phase of tail movements dynamically adjusted based on their front-back spatial separation (static water: Fig. 1J, swim tunnel: Fig. 1K), a phenomenon known as "vortex phase matching" [44]. This vortex phase matching was absent in session-shuffled null models of these data (Fig. 1, J and K).

The occurrence of vortex phase matching suggests the existence of a circuit mechanism for the online integration of allocentric signals (vortex-generated body deflections) to dynamically coordinate and synchronize fish body movements within the group. A potential candidate for detecting body bending are the centrally located proprioceptors, specialized stretch-sensitive glycinergic neurons located along the internal edges of the spinal cord [31] (Fig. 2A). Using calcium imaging in Tg(*glyt2*-GCaMP6s) transgenic zebrafish, we observed that these intraspinal proprioceptors exhibited a clear increase in activity in response to imposed body bending (Fig. 2, B and C). These neurons project via contralateral ascending axons to form direct inhibitory connections with locomotor rhythm-generating V2a interneurons [31]. Consistent with this circuit configuration, calcium imaging of V2a interneurons using Tg(*chx10*-GCaMP6s) transgenic zebrafish revealed a corresponding decrease in calcium signal during bending (Fig. 2, B and C).

**Self-generated proprioceptive feedback dictates locomotor timing**

We next sought to determine whether the proprioceptor-V2a interneuron circuit can functionally integrate body curvature feedback to regulate locomotor dynamics. To test this, we used Tg(*vglut2*-ChR2) transgenic zebrafish to induce swimming in head-fixed preparations via optogenetic stimulation of glutamatergic brainstem projection neurons. Activity was monitored using EMG recordings while left-right tail bending was unrestricted (proprioception ON; Fig. 2, D to F) or restricted (proprioception OFF; Fig. 2, G to I).

In the proprioception OFF condition, where self-generated (egocentric) feedback was absent, the frequency of locomotor bursts was significantly reduced compared to proprioception ON (Fig. 2, D, G and J). A similar decrease in frequency was also directly recorded in V2a interneurons, which was associated with a delayed termination of the V2a excitation window (Fig. 2, E and H, left traces). In freely bending conditions (proprioception ON), the excitation window of V2a interneurons terminated sharply, whereas in the absence of bending feedback, this window was significantly prolonged (Fig. 2, E and H, right traces and Fig. 2K). These effects were mirrored by changes in the inhibitory currents, which were significantly larger when tail bending was unrestricted (Fig. 2, F, I and L). This shows that phasic, sensory-mediated inhibitory feedback resulting from tail bending is responsible for terminating rhythmic V2a excitation, thereby shortening the locomotor cycle and increasing speed.

Together these results indicate that phasic proprioceptive feedback from tail movements sets the timing of the rhythm-generating V2a interneurons, and consequently the overall locomotor central pattern generator (CPG). These findings align with predictions from modeling and robotics studies showing that curvature-based feedback is essential for locomotor speed and stability, and they highlight proprioception as a central mechanism in movement coordination [45-49]. Thus, the minimal sensorimotor circuit formed by spinal proprioceptors and V2a interneurons can detect and integrate egocentric feedback to coordinate and sustain efficient swimming.

**A low-order sensorimotor circuit for locomotor adaptation**

One advantage of locomotor timing being regulated by sensory feedback is that it would allow the locomotor CPG to automatically adjust to dynamic or unpredictable changes in the environment. To test whether this circuit can also integrate allocentric signals from the external environment, we imposed a static tail bend to mimic a constant sideways force during swimming. This stretches the spinal cord on one side, tonically activating the intraspinal proprioceptors. Both the calcium activity of V2a interneurons and the swim motor output became asymmetrical across the left and right sides. Whole-cell recordings showed that rhythmic V2a activity contralateral to the bend was strongly suppressed by direct tonic inhibition, reducing rhythmic excitation and firing in motor neurons, whereas ipsilateral motor neurons and some V2a interneurons were disinhibited and increased their activity. V2a interneurons are therefore at the core of a circuit that integrates proprioceptive feedback in response to external forces and propagates it throughout the locomotor CPG.

To test whether rhythmic allocentric signals can pace the locomotor CPG, we first imposed symmetrical tail bends at fixed frequencies during swimming induced by brainstem optogenetic stimulation in *ex vivo* preparations (*ex vivo* pacing, Fig. 3A). Swim motor bursts became tightly aligned to the pacing signal, synchronizing swim frequency to that of the imposed tail bending (Fig. 3, B and C). During natural swimming, however, the frequency of tail movements is not fixed but varies dynamically, with this variability reflected in the properties of the vortex trails within a school. To mimic this using *ex vivo* preparations, we imposed a sequence of dynamic tail movements derived from a leading fish in a natural schooling interaction, i.e. with time-varying frequencies (*ex vivo* schooling, Fig. 3D). In the absence of CPG activity (no swimming), these schooling signals produced a time-locked sequence of inhibition in V2a interneurons (Fig. 3E). During swimming, the firing of V2a interneurons, and consequently the entire resulting motor output, directly followed the timing of the schooling signals (Fig. 3E). These results show that proprioceptor-driven inhibition in V2a interneurons can dynamically reset and synchronize rhythm generation, providing a mechanism for the exploitation of complex allocentric signals during schooling.

***In silico* neuromechanical model for sensorimotor integration**

To assess whether these circuit mechanisms are sufficient to drive the inter-individual swim synchronization observed within a school, we implemented our findings in a spiking neuromechanical model of the adult zebrafish locomotor circuit. The network incorporates the major classes of neurons critical to rhythm generation and sensorimotor integration, including excitatory V2a interneurons, inhibitory V0d interneurons, motor neurons, and intraspinal proprioceptive neurons. Neural properties, spiking patterns, and connectivity were constrained based on available experimental data across several studies in adult zebrafish (see Materials and Methods and Supplementary Material) [27,50-52]. To simulate physical interaction with the environment, the network was bidirectionally coupled to a mechanical model of a zebrafish body with articulated joints actuated by simulated muscle models, implemented within the MuJoCo physics engine [36]. Crucially, stretch-sensitive proprioceptive neurons were integrated into the model network to detect local body curvature and transmit ascending inhibitory signals selectively targeting V2a interneurons [31]. This model produced appropriately coordinated swimming activity.

We first used this model to test the impact of a loss of curvature-based proprioceptive feedback on self-driven swimming *in silico*. Mirroring our experiments in real fish, the absence of proprioceptive feedback (proprioception OFF) resulted in a reduction in the frequency of swimming activity compared to proprioception ON. This frequency change was mediated by a lengthening of the excitation window in modeled V2a interneurons resulting from a decrease in inhibition. Similarly, the *in silico* model recapitulated all motor output responses to externally imposed tail movements. When rhythmic tail bending was applied to the model, the neural network exhibited clear pacing behavior (*in silico* pacing, Fig. 3, F to H). Specifically, imposed rhythmic bending successfully synchronized the neural network's ongoing oscillations, adjusting the frequency of the locomotor rhythm to a higher or lower frequency depending on the pacing signal.

The neural network's activity could also synchronize robustly with imposed bending signals of dynamically varying frequencies, derived from kinematic patterns of leader-follower interactions observed during *in vivo* schooling. Dynamic rhythmic bending elicited synchronized inhibition of V2a interneurons during network quiescence and precisely timed rhythmic spiking when the network was actively oscillating (*in silico* schooling, Fig. 3, I and J).

**Proprioception-driven hydrodynamic coupling in silico and in robot**

To test if our model has the capacity to adapt to naturalistic hydrodynamic perturbations generated by neighboring model zebrafish, we built a fluid model environment (see Materials and Methods). In this configuration the leader model fish generates a vortex trail that is sufficient to influence the body of the follower model fish. The follower fish, controlled by the neural network, is immersed in the wake of the leader and interacts with the incoming turbulent flow. Under these conditions, the two simulated fish synchronized their activity (*in silico* vortex matching), with the swim frequency of the follower (blue) model fish becoming aligned with that of the leader (yellow, Fig. 4, A and B). In the presence of proprioceptive feedback (proprioception ON, closed loop), the relative phase between leader and follower dynamically adjusted as a function of their relative distance, the hallmark of vortex matching during schooling (Fig. 4, B and C). When proprioceptive feedback was removed from the model (proprioception OFF, open loop), there was a loss of both network synchronization and vortex matching (Fig. 4, D and E). Under these conditions, there was no relationship between the relative phase and distance between the leader and follower model fish (Fig. 4, E and F). Vortex phase matching emerged only in the closed loop condition across a broad range of muscle stiffness, including the values used here, and was lost when the body was made either highly compliant or highly stiff (see Supplementary Material).

We next sought to confirm the performance of the model in a real fluid environment with real-world embodiment. For this, we developed a bio-inspired fish robot driven by the circuit model of adult zebrafish. The robot fish was placed in a water channel where vortex trails were generated using a controllable flapper to mimic leader fish tail movements during swimming. In the presence of proprioceptive feedback (proprioception ON, closed loop), the robot synchronized its swimming with the movements of the flapper (Fig. 4, G and H). When the robot was placed at different distances from the flapper, it dynamically adjusted the phase of its swimming movement relative to that of the flapper (Fig. 4I). In the absence of proprioceptive feedback (proprioceptor OFF, open loop), the robot was no longer able to produce vortex matching as there was no relationship between the relative phase and distance between the robot and flapper (Fig. 4, J and K).

Finally, we examined whether vortex matching provides additional adaptive energetic benefits. For this, we quantified energy consumption in the presence and absence of vortices both *in silico* (Fig. 4, L and M) and in the robot (Fig. 4, N and O). In the proprioception ON condition (closed loop) there was a significant decrease in energy consumption in the presence of vortices, both *in silico* (Fig. 4L) and in the robot (Fig. 4N). In the proprioception OFF condition (open loop) there was no energy benefit in the presence of vortices either *in silico* (Fig. 4M) or in the robot (Fig. 4O). These results demonstrate that proprioception-enabled vortex matching confers an energetic advantage during schooling, in line with previous studies [44,53,54]. Thus, the implementation of a single, appropriately configured source of proprioceptive feedback in our high-fidelity model and robot equips the network with adaptability and the ability to encode hydrodynamic perturbations from conspecifics while reducing energetic cost.

**Proprioception-driven collective schooling dynamics**

The results from our model and robot align with the *in vivo* vortex matching observed experimentally in wild-type zebrafish (Fig. 5, A to C, see also Fig. 1, J and K). Adding this feedback loop led to vortex matching, and removing it abolished it. To test this prediction in freely swimming fish, we disrupted this proprioceptive feedback by introducing a loss-of-function mutation in the gene encoding the mechanosensitive ion channel Piezo2 expressed by intraspinal proprioceptors [31]. This mutation resulted in a significant decrease in *piezo2* expression in intraspinal proprioceptors. Using whole-cell patch clamp recordings combined with mechanoclamp stimulation, we confirmed that this was associated with a decrease in the firing and the amplitude of the mechanosensitive current in response to applied mechanical force.

As predicted by our modelling and robotics experiments, there was a loss of vortex phase matching in *piezo2* mutant zebrafish, with no clear relationship between the relative swim phase and separation distance (Fig. 5, D to F). Analysis of *piezo2* mutants at the group level revealed that they no longer school (*in vivo* schooling, Fig. 5G) and failed to form tight, cohesive groups within the swim tunnel (Fig. 5H). Overall, mutant fish had a significantly larger polygonal area compared to wild-type fish (Fig. 5I). A similar loss of cohesive group swimming in the swim tunnel was observed in a group of fish in which intraspinal proprioceptors were selectively eliminated using targeted two-photon ablation.

Next, we tested the behavior of *piezo2* mutant zebrafish in static water conditions. Mutant fish failed to form organized groups, with a loss of polarization (Fig. 5, J and K). To further support this finding and mitigate possible broader effects of *piezo2* mutation, we used intersectional genetics to eliminate intraspinal proprioceptors through selective DTA expression. The elimination of intraspinal proprioceptors abolished collective schooling, with a complete loss of group polarization in static water (Fig. 5, J and K). Overall, the group polarization index was significantly lower in mutant and DTA-ablated fish compared to wild-type fish (Fig. 5L).

Together, our computational modeling, robotics, and loss-of-function experiments demonstrate that a localized, low-order spinal sensorimotor circuit underpins vortex matching, enabling the real-time collective synchronization essential for schooling behavior.

**Discussion**

Social behaviors are widely considered to ultimately emerge from high-level cognitive processing in the brain [2,3,5,6,11-13]. Here, by combining experimental, computational and robotics approaches, we reveal a low-order spinal circuit that is required for vortex phase matching, underlying the coordination necessary for collective schooling behavior. Central to this circuit are intraspinal proprioceptors, which are present in other aquatic vertebrates, where they also have a role in entraining locomotion in response to body bending [55-61]. These neurons sense tension changes at the edges of the spinal cord during body movements and deliver real-time, curvature-based feedback. In zebrafish, by providing selective and direct inhibition to the core rhythm generating V2a interneurons, this rapid proprioceptive feedback dictates the timing of the locomotor CPG. The power of this low-level sensorimotor circuit is that it autonomously paces the locomotor CPG according to any tail deflection – static or dynamic, self-generated or allocentric – including those generated by conspecific vortex trails during schooling.

Zebrafish exhibit a wide repertoire of social behaviors including shoaling, schooling, social preference, and aggression [62-69]. Whereas shoaling refers broadly to the tendency to remain in close proximity, schooling is distinguished by polarized swimming and collective coordination within the group [70], and confers advantages such as energy saving, protection from predation, and improved foraging success [10,14,44,53,71-74]. Its emergence is thought to involve complex temporal dynamics [75] and multimodal sensory integration, with group formation, orientation and threat detection engaging a range of sensory modalities including vision, olfaction, hearing and lateral line flow detection [13,76].

Central to the emergence of collective, orchestrated motion within fish schools is the rapid integration of local information flows governed by simple behavioral rules [16]. Early studies described the significance of a close spatial organization of individuals and specific formation patterns within schools [73,77]. These formations are considered to provide an optimal hydrodynamic architecture to exploit vortex trails from immediate neighbors and align their behavior for energy saving ("vortex phase matching") [44,71,76,78]. The sensory system detecting this local information flow within the school, and how this is integrated within central circuits, has remained elusive. The roles of vision and the lateral line have been tested in previous studies, with effects of a loss of these sensory modalities ranging from no detectable change to altered cohesion and polarization, differing in direction and magnitude between species [44,79-82]. A proprioceptive basis for this behavior has previously been suggested [44], but the underlying sensor and circuit have remained unidentified.

Here, we have identified a proprioceptive pathway acting within a parsimonious spinal circuit as the key sensory mechanism for detecting and integrating local hydrodynamic information necessary for schooling behavior. In the absence of proprioceptive feedback activated by body bending, vortex phase matching is lost and zebrafish no longer form polarized, cohesive schools. While the disruption of this circuit in *piezo2* mutants affects their ability to sustain high-frequency swimming, as reported previously [31], this locomotor change does not account for their impaired social coordination since *piezo2* mutants fail to school across their entire range of swimming speeds, even at swim speeds at which wild-type fish school normally. In our model and robot, where this feedback loop is the only source of coupling between the control circuits of individuals, its presence is sufficient to generate vortex phase matching and to reduce the energetic cost of swimming in the wake of a neighbor. Our results therefore show that the continuous, local integration of hydrodynamic forces by this spinal circuit is required for the moment-to-moment phase coordination between neighbors. In the behaving animal, complex descending inputs integrating multiple sensory modalities will likely interact with this spinal circuit to regulate schooling. The core task of online phase coordination can be delegated to this low-order schooling control system, allowing slower, high-order systems in the brain to finesse behavior and render it context-dependent.

Collective motion is a spectacular manifestation of coordinated social behavior that exists across all scales of the animal kingdom [83]. Self-organization theory has been proposed as a unifying framework for understanding the basic principles of collective motion, suggesting that global patterns in biological systems can emerge spontaneously through simple, local interactions governed by low-order processing rules [16]. This theory has been widely explored through mathematical models, swarm robotics and computational simulations [84-88], drawing parallels with Conway's Game of Life [89]. In biological systems, however, the nature of the rule and its implementation have remained unknown. Our study now uncovers both the sensor and the circuit that implement such a rule for synchronization during schooling. We show that simple, rule-based local interactions couple individuals to their neighbors, allowing collective behavior to emerge without centralized control.

## Figure Legends

**Figure 1.** Schooling behavior and vortex matching in zebrafish. **A,** Schematic of schooling zebrafish with a leader (yellow) and follower fish (blue). Blue and red trails represent leader-generated vortices. **B,** Schematic of schooling fish modeled as a chain of oscillators integrating both egocentric (self-generated) and allocentric (externally generated) sensory cues. **C,** Top: experimental setup for *in vivo* schooling in flow tunnel. Bottom: representative frame of schooling zebrafish overlaid with polygonal pattern of zebrafish positions within the school. **D,** Control fish (green) formed tight group constellations in the swim tunnel represented by small polygonal area, which was significantly smaller than polygonal area in a shuffled null model (grey) ($p = 0.0117$, Wilcoxon signed-rank test, one-sided, $n = 8$ batches, 1000 shuffles). **E,** Representative timepoints showing groups of 8 control zebrafish swimming in a circular dish under static water conditions. Fish are represented as arrows indicating their position and swimming direction. **F,** Control zebrafish spontaneously form highly polarized groups compared with shuffled null distributions. **G,** Quantification of group polarization showing significantly higher polarization in control zebrafish compared with a shuffled null model ($p = 0.002$, Wilcoxon signed-rank test, one-sided, $n = 9$ batches, 1000 shuffles). **H,** Leader-follower interactions between zebrafish ($\Psi$: tail angle). **I,** Tail angles and phase synchronization of leader and follower tail movements during swimming. **J,** Vortex matching in static water: relative phase ($\Delta\Phi$) plotted as a function of inter-fish distance (control $R^2 = 0.571$, $p = 0.0138$, $n = 9$ batches, linear-circular correlation analysis). Permutation testing showed a significant difference between control zebrafish and a shuffled null model ($p = 0.033$, $n = 9$ batches, $R^2$ permutation test, 1000 shuffles). **K,** Vortex matching in flow tunnel: relative phase ($\Delta\Phi$) plotted as a function of inter-fish distance (control $R^2 = 0.548$, $p = 0.00718$, $n = 10$ pairs; linear-circular correlation). Permutation testing of $R^2$ values

revealed significant differences between control fish and a shuffled null model (p = 0.031, n = 10 pairs, $R^2$ permutation test, 1000 shuffles).

**Figure 2.** Egocentric proprioceptive feedback sculpts locomotor timing. **A,** Spinal sensorimotor circuit encoding tail bend (red: V2a interneurons, yellow: motor neurons, green: intraspinal proprioceptors). **B,C,** Spinal bend increased calcium activity in intraspinal proprioceptors (****p < 0.0001, n = 5 cells from 2 fish; paired two-tailed t-test) and decreased calcium activity in contralateral V2a interneurons (***p = 0.0005, n = 6 cells from 3 fish; paired two-tailed t-test). **D,** Left: experimental setup with free tail movement (proprioception ON), with skin removed for access of EMG recording electrode. Right: EMG-recorded swimming activity in tail free configuration. **E,** Activity recorded in a V2a interneuron (left) with several cycles of excitation superimposed (right) in the presence of proprioceptive feedback. **F,** Outward inhibitory currents recorded in a V2a interneuron during swimming with tail free. **G,** Left: experimental setup with restricted tail movement (proprioception OFF). Right: EMG recorded swimming activity in tail blocked configuration. Faded blue dots show tail-free swim frequencies. **H,** Activity recorded in a V2a interneuron (left) with several cycles of excitation superimposed (right) in the absence of proprioceptive feedback. For this, we used *ex vivo* preparation where skin and muscle at the recording site was removed. **I,** Outward inhibitory currents recorded in a V2a during swimming with tail blocked. **J-L,** Graphs showing pooled data of swim frequency (**J**, p = 0.0013, 9 cells from 5 fish; paired two-tailed t-test), on-cycle excitation (**K**, p < 0.0001, 9 cells from 5 fish; paired two-tailed t-test) and amplitude of inhibitory currents (**L**, p = 0.0041, 8 cells from 5 fish; paired two-tailed t-test) in proprioception ON (blue) and proprioception OFF (magenta).

**Figure 3.** Allocentric signals pace locomotion *ex vivo* and *in silico***.** **A,** Experimental setup for *ex vivo* pacing. **B,** EMG recording of swimming activity during pacing at 2 Hz (left) and 8 Hz (right). **C,** Frequency of swimming before, during and after pacing at different frequencies (2, 4, 6, 8, and 10 Hz). There was a strong linear correlation between the pacing frequencies and swim frequencies ($R^2 = 0.9964$, n = 10 fish; linear regression). **D,** Experimental setup for *ex vivo* schooling**. E,** In the absence of swimming, the schooling signal induced time-locked inhibition in V2a interneurons. During swimming, the schooling signal synchronized the firing of V2a interneurons and EMG swim bursts (pre vs. pacing: $p < 0.0001$, pacing vs. post: $p < 0.0001$, n = 8 fish; repeated measures one-way ANOVA followed by Tukey's post-hoc test for multiple comparisons). Heatmap represents the normalized synchronization of swim bursts (synchronization index: SI), with values of 1 and -1 representing maximum synchronization for active and inactive phases, respectively. **F,** Neuromechanical model for *in silico* pacing. **G,** Simulated swimming during pacing at 2 Hz (left) and 8 Hz (right). **H,** Frequency of the simulated swimming before, during and after pacing at different frequencies. There was a strong linear correlation between the pacing frequencies and swim frequencies ($R^2 = 0.9998$, n = 3 networks tested; linear regression). **I,** Neuromechanical model for *in silico* schooling. **J,** Schooling signal induced rhythmic inhibition in V2a interneurons in the silent network and synchronized both V2a interneurons rhythmic firing and simulated muscle activity in an actively oscillating network (pre vs. pacing: $p < 0.0001$, pacing vs. post: $p=0.0003$, n = 15 networks; repeated measures one-way ANOVA followed by Tukey's post-hoc test for multiple comparisons). Heatmap represents the normalized synchronization of swim bursts (synchronization index: SI), with values of 1 and -1 representing maximum synchronization for active and inactive phases, respectively.

**Figure 4.** *In silico* and robotic vortex matching and energy saving. **A,** Simulation of leader–follower interactions within a fluid dynamics model showing *in silico* vortex matching. **B,** Time series of body angles and swimming phases for leader (yellow) and follower (blue) simulated fish, demonstrating phase synchronization. **C,** Swimming interactions as a function of phase difference ($\Delta\Phi$) and front-back distance in simulation. Follower fish adjust their swimming phase according to their relative distance to the simulated leader ($R^2 = 0.558$, n = 5 networks; explained circular variance). **D-F,** In the proprioceptor-OFF simulation, the follower fish decoupled from the leader and failed to perform vortex matching ($R^2 = 0.074$, n = 5 networks; explained circular variance; permutation testing of $R^2$ values: proprioceptor-ON vs. proprioceptor-OFF, $p < 0.0001$). **G,** Single images showing robotic fish ("follower robot", blue) and flapper (yellow) at three timepoints showing real-world vortex matching in the proprioceptor-ON condition. **H,** Time series of angles and swimming phases for flapper (yellow) and robot (blue), demonstrating phase synchronization in the proprioceptor-ON condition. **I,** Flapper-robot interactions plotted as a function of phase difference ($\Delta\Phi$) and front-back distance. The robotic fish adjusted its swimming phase according to its relative distance to the flapper ($R^2 = 0.711$, n = 6 trials; explained circular variance). **J,K,** In the proprioceptor-OFF condition, the robot decoupled from the flapper and failed to perform vortex matching ($R^2 = 0.035$, n = 6 trials; explained circular variance; permutation testing of $R^2$ values: proprioceptor-ON vs. proprioceptor-OFF, $p < 0.0001$). **L,M,** Power consumption was significantly lower in the presence of leader-generated vortices compared to solo swimming in the proprioceptor-ON simulation (~17% decrease, $p < 0.0001$, n = 50 networks; two-tailed unpaired t-test), with no differences in the proprioceptor-OFF simulation ($p = 0.06$, n = 50 networks; two-tailed unpaired t-test). **N,O,** Power consumption of the robot was significantly lower in the presence of flapper-generated vortices compared to solo swimming in the proprioceptor-ON condition (~7% decrease, $p < 0.0001$, n = 6 trials; two-tailed

unpaired t-test) but not in the proprioceptor-OFF condition ($p = 0.1061$, $n = 6$ trials; two-tailed unpaired t-test).

**Figure 5.** Proprioceptive-driven vortex matching and cohesive schooling behavior *in vivo*. **A,** Frames from video recordings of wild-type zebrafish showing tail position matching during close interactions *in vivo*. **B,** Time series of leader and follower angles showing *in vivo* vortex phase matching. **C,** Wild-type vortex matching in flow tunnel: relative phase ($\Delta\Phi$) plotted as a function of inter-fish distance ($R^2 = 0.548$, $p = 0.00718$, $n = 10$ pairs; linear-circular correlation). **D,** Frames from video recordings of *piezo2* mutant fish show tail position mismatches during close interactions. **E,** Time series of leader and follower angles showing unpaired phases. **F,** *Piezo2* mutant fish are unable to perform *in vivo* vortex matching in flow tunnel. Relative phase ($\Delta\Phi$) plotted as a function of inter-fish distance ($R^2 = 0.141$, $p = 0.281$, $n = 6$ pairs; linear-circular correlation). Permutation testing of $R^2$ values revealed significant differences between control fish and *piezo2* mutant fish ($p = 0.001$). **G,** Representative frame from video recording of *piezo2* mutant fish showing large dispersion in the tank and the corresponding polygonal pattern. **H,** Observed polygonal pattern of *piezo2* mutant fish positions (red; $n = 8$ batches, 40 frames overlaid) compared to wild-type fish (green; $n = 8$ batches, 40 frames overlaid). **I,** Quantification of normalized polygonal area across conditions. *Piezo2* mutant zebrafish showed a loss of group stability and cohesion as reflected in a significantly larger polygonal area compared to wild-type fish (wild-type: $n = 8$ batches; *piezo2* mutant: $n = 8$ batches, two-tailed unpaired t-test; ****$p < 0.0001$). **J,** Representative snapshots at three timepoints of groups of wild-type, *piezo2* mutant, and conditionally ablated zebrafish (DTA ablation) in static water conditions. **K,** Comparison of the distribution of group polarization values in wild-type, mutant and DTA-ablated fish. **L,** Comparison of group polarization values across conditions. *Piezo2* mutants and DTA-ablated

fish showed significantly lower polarization compared to wild-type fish (wild-type: n = 9 batches; genetic ablation: n = 5 batches; *piezo2* mutant: n = 8 batches, One-way ANOVA followed by Tukey's post-hoc test for multiple comparisons; ** $p < 0.01$).

# Figures

## Figure 1

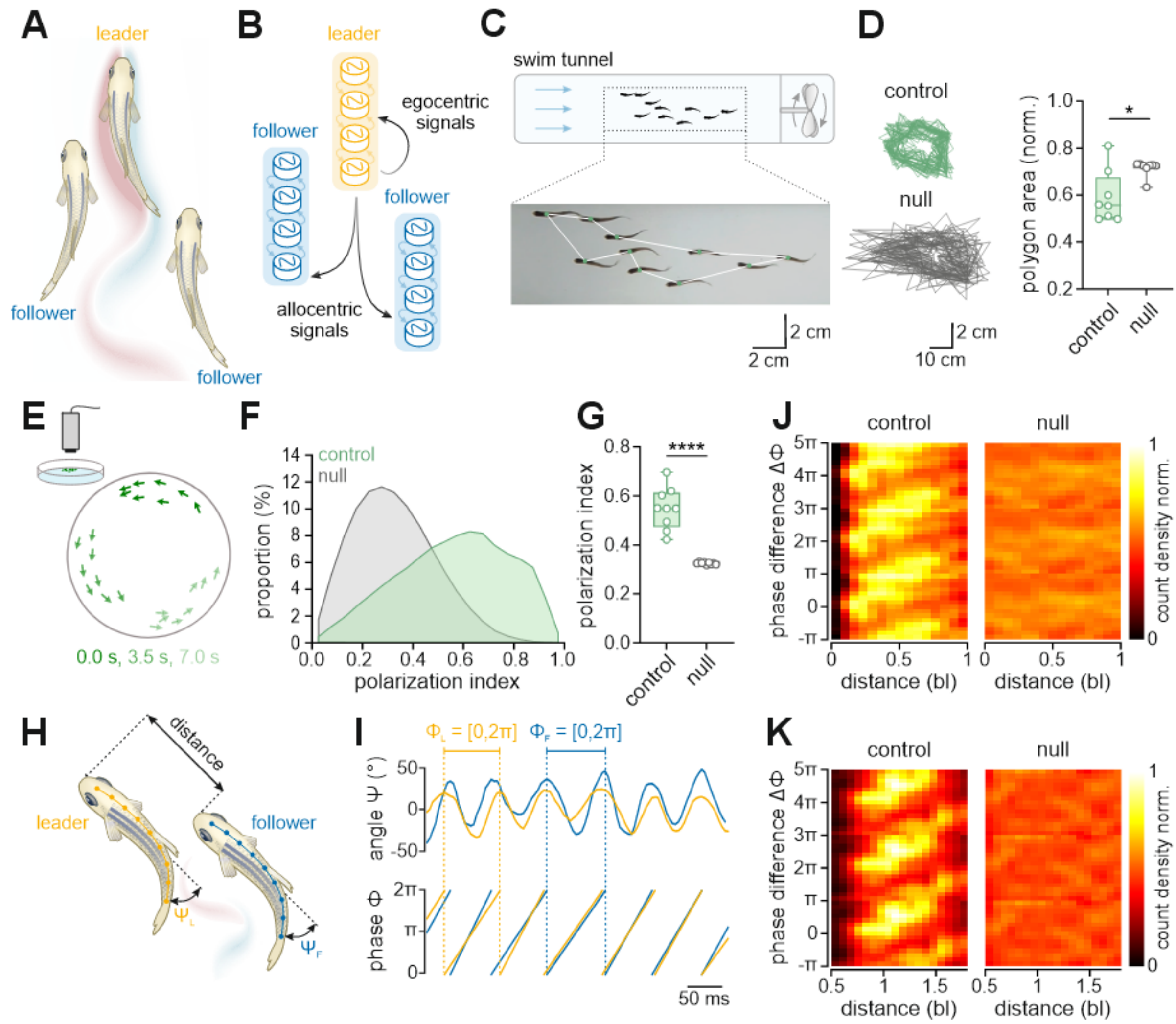

Figure 2

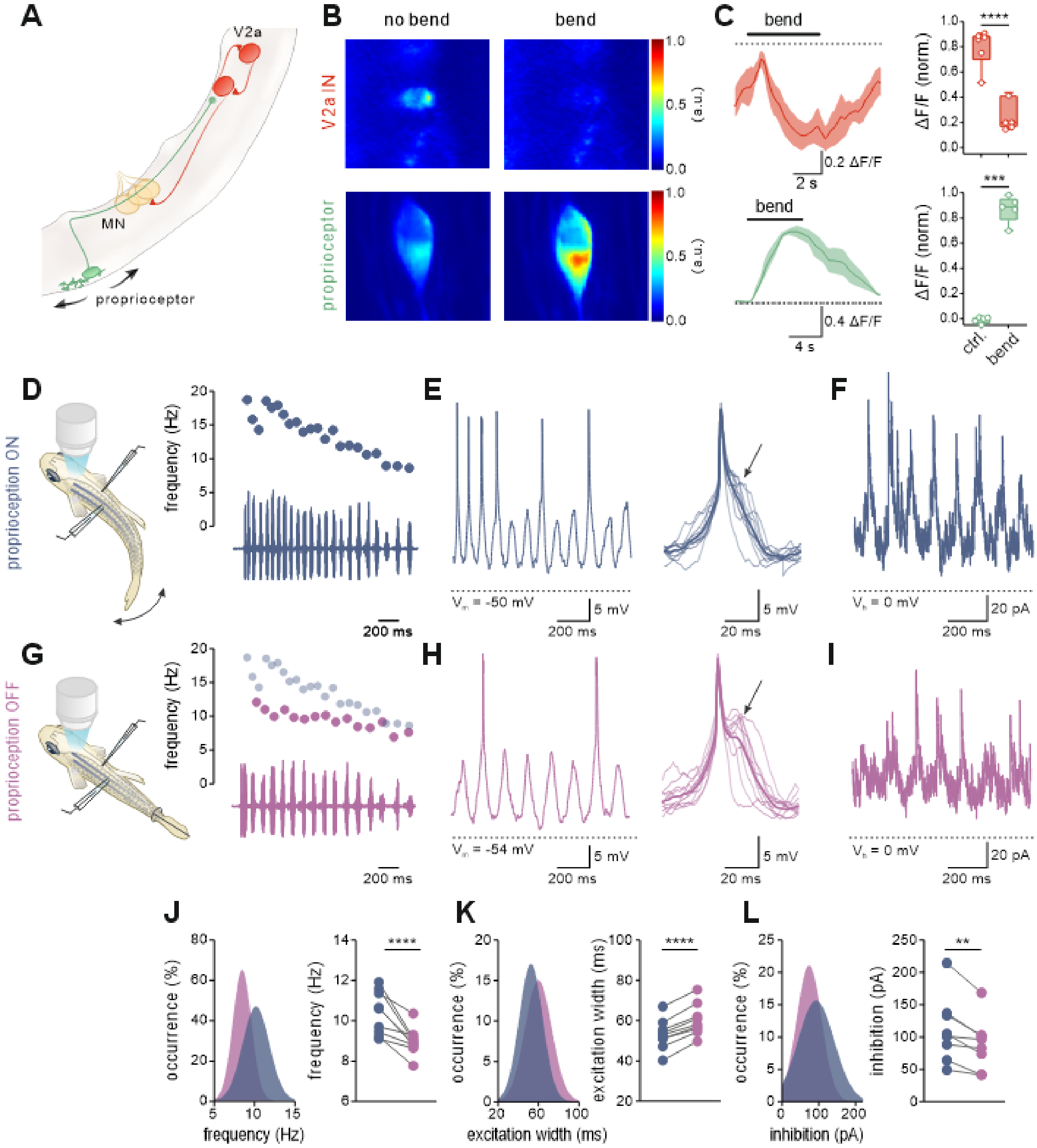

A
V2a
MN
proprioceptor
B
no bend
bend
V2a IN
proprioceptor
1.0
0.5
0.0
(a.u.)
C
bend
0.2 ΔF/F
2 s
0.4 ΔF/F
4 s
ΔF/F (norm.)
****
***
ctrl.
bend
D
proprioception ON
frequency (Hz)
200 ms
E
F
$V_m$ = -50 mV
5 mV
200 ms
20 ms
$V_h$ = 0 mV
20 pA
G
proprioception OFF
H
I
$V_m$ = -54 mV
J
occurrence (%)
frequency (Hz)
K
excitation width (ms)
L
inhibition (pA)
**

**Figure 3**

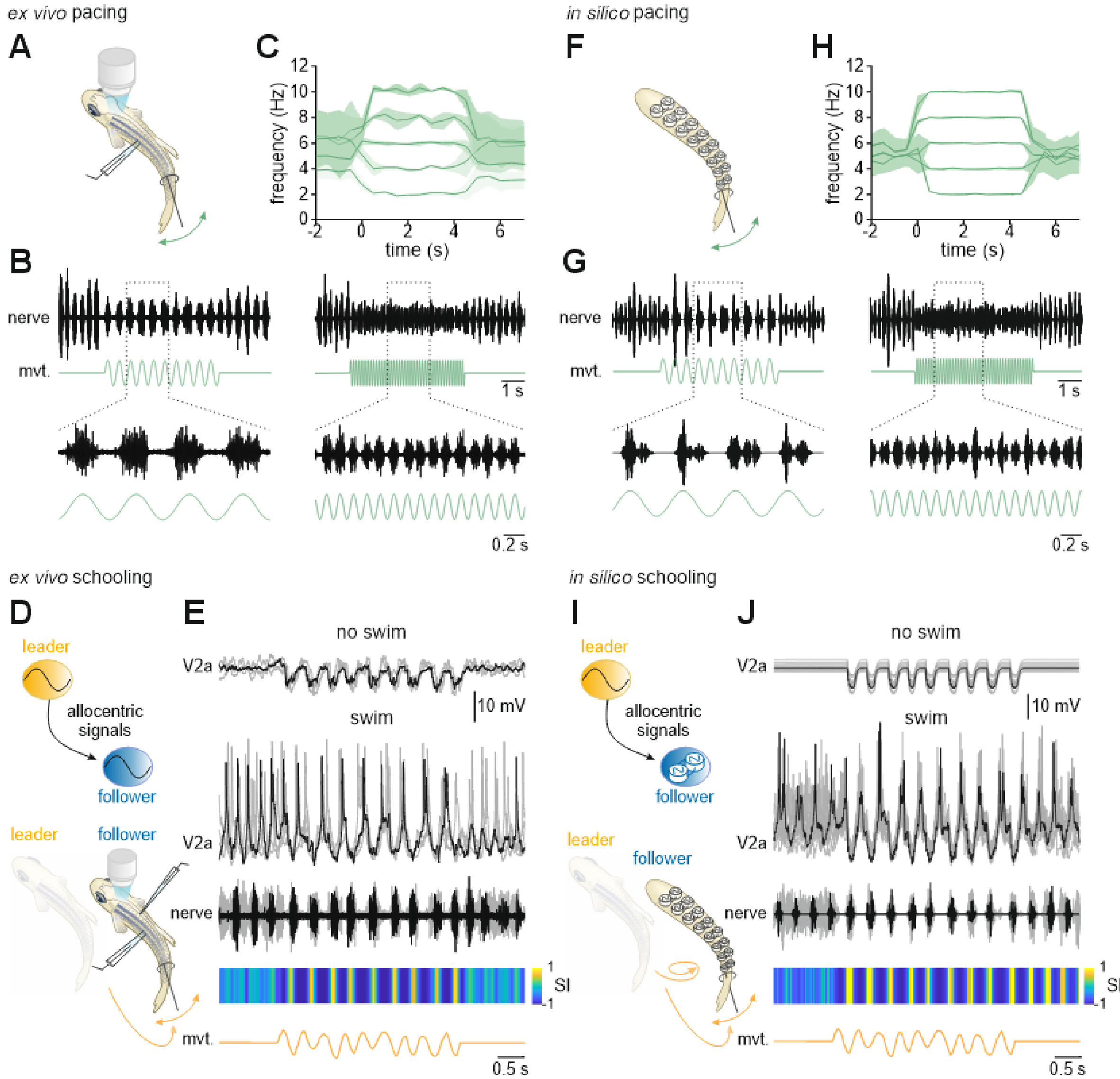

ex vivo pacing
A
C
frequency (Hz)
time (s)
B
nerve
mvt.
1 s
0.2 s
in silico pacing
F
H
frequency (Hz)
time (s)
G
nerve
mvt.
1 s
0.2 s
ex vivo schooling
D
leader
allocentric signals
follower
leader
follower
E
no swim
V2a
10 mV
swim
V2a
nerve
SI
mvt.
0.5 s
in silico schooling
I
leader
allocentric signals
follower
leader
follower
J
no swim
V2a
10 mV
swim
V2a
nerve
SI
mvt.
0.5 s

**Figure 4**

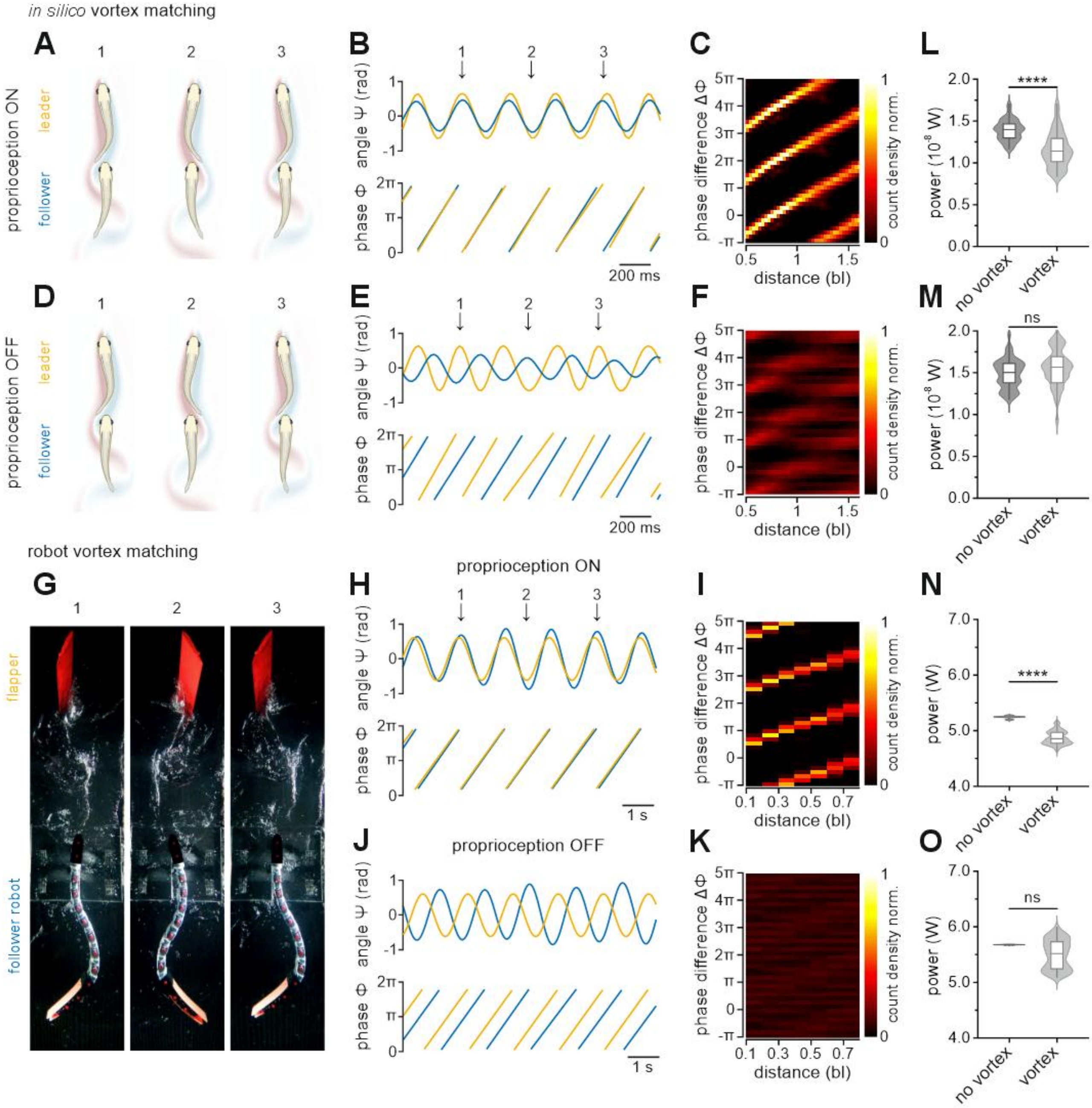

in silico vortex matching
A
B
C
L
proprioception ON
leader
follower
angle Ψ (rad)
phase Φ
200 ms
phase difference ΔΦ
count density norm.
distance (bl)
power (10⁻⁸ W)
no vortex
vortex
****
D
E
F
M
proprioception OFF
ns
robot vortex matching
G
H
I
N
flapper
follower robot
proprioception ON
1 s
J
K
O
proprioception OFF
power (W)

**Figure 5**

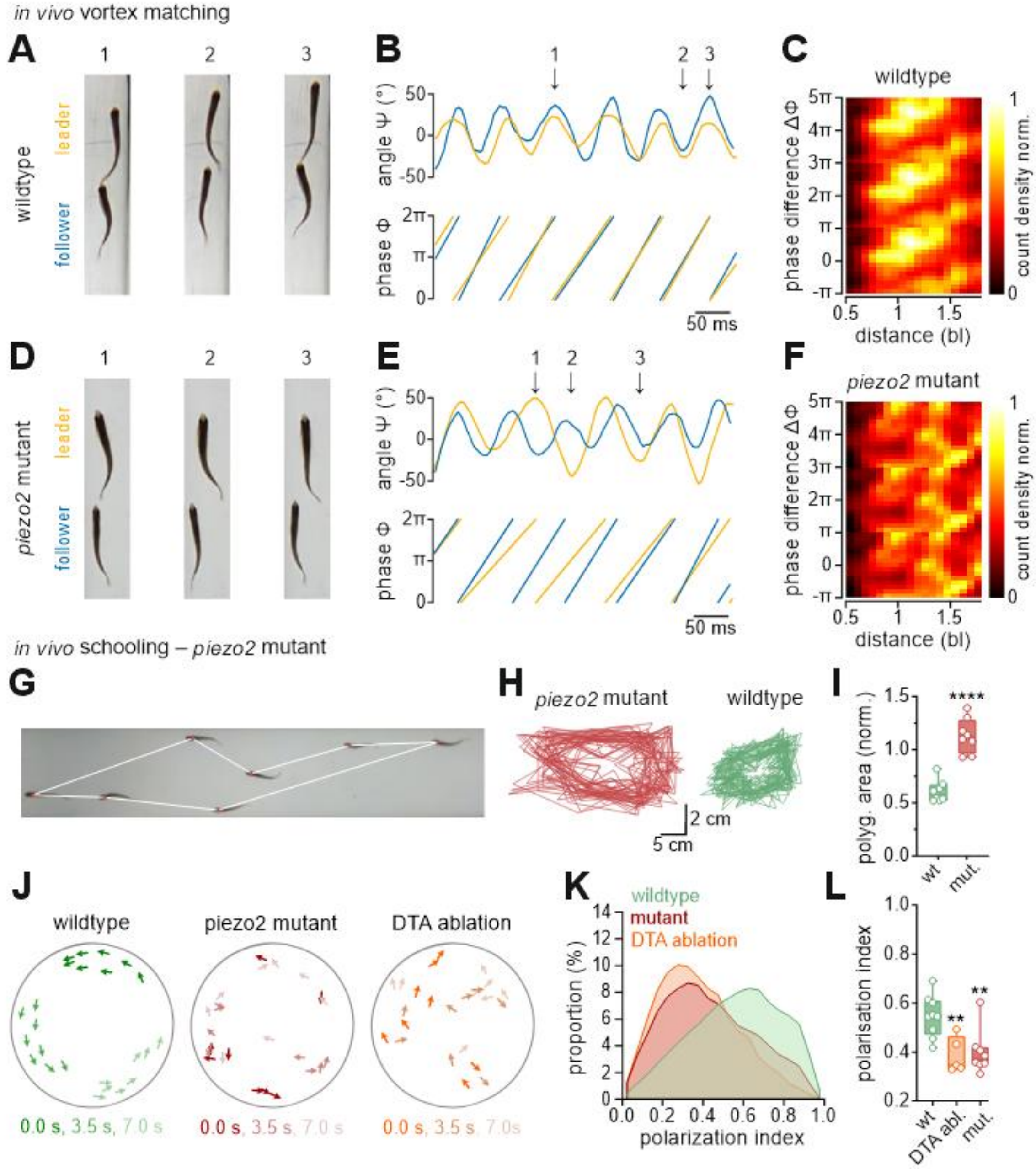

in vivo vortex matching
A
1
2
3
wildtype
leader
follower
B
angle Ψ (°)
phase Φ
50 ms
C
wildtype
phase difference ΔΦ
count density norm.
distance (bl)
D
piezo2 mutant
leader
follower
E
F
piezo2 mutant
in vivo schooling – piezo2 mutant
G
H
piezo2 mutant
wildtype
2 cm
5 cm
I
polyg. area (norm.)
wt
mut.
J
wildtype
piezo2 mutant
DTA ablation
0.0 s, 3.5 s, 7.0 s
K
wildtype
mutant
DTA ablation
proportion (%)
polarization index
L
polarisation index
wt
DTA abl.
mut.

# Supplementary Materials

## 1) Neuronal model

**Sub-threshold dynamics:** In this model, neurons are modeled as adaptive exponential integrate-and-fire units [1]. The subthreshold dynamics of the membrane potential $u$ and adaptation variable $\omega$ are governed by:

$$\tau_m \frac{du}{dt} = -(u - E_{rest}) + \Delta_T \exp\left(\frac{u - E_{rh}}{\Delta_T}\right) - R_m \omega + R_m I_{syn} + R_m I_{ext} \tag{1}$$

$$\tau_\omega \frac{d\omega}{dt} = a_\omega (u - E_{rest}) - \omega \tag{2}$$

In this context, $E_{rest}$ is the neuronal resting potential, $E_{rh}$ is the rheobase threshold, $\Delta_T$ is the slope factor controlling the sharpness of spike initiation, $R_m$ is the input membrane resistance and $\tau_m$ is the membrane time constant. The evolution of the adaptation variable is governed by the time constant $\tau_\omega$ and by the parameter $a_\omega$ controlling the subthreshold adaptation. The input to the model comes from the total synaptic current $I_{syn}$ and the external input current $I_{ext}$.

**Spike Detection and Reset Dynamics:** When the membrane potential reaches the firing threshold $E_{thres}$, a spike is detected and the following reset conditions are applied:

$$\text{if } u(t^{f-}) \geq E_{thres}: \qquad \begin{array}{c} u(t^{f+}) \leftarrow E_{reset} \\ \omega(t^{f+}) \leftarrow \omega(t^{f-}) + \Delta\omega \end{array} \tag{3}$$

where $t^{f-}$ and $t^{f+}$ denote the time immediately before and after the spike, respectively, $E_{reset}$ is the reset voltage and $\Delta\omega$ is the increment for the adaptation variable.

**Refractory Period:** Following each spike, the neuron enters a refractory period of duration $t_{refr}$ during which the membrane potential is clamped:

$$u(t) = E_{reset}, \quad \text{for } t \in [t^{(f)}, t^{(f)} + t_{refr}] \tag{4}$$

**Synaptic Dynamics:** The total synaptic current is given by the contribution of different synaptic conductances $g_s$ and their respective synaptic reverse potential $E_s$ according to

$$I_{syn} = \sum_s g_s (E_s - u) \tag{5}$$

The conductance $g_s$ sees an increment by $\Delta g_s$ with every incoming spike [2]. Additionally, there is a constant synaptic transmission delay of 2ms between the spike event and the corresponding $g_s$ modification. The temporal evolution of synaptic conductances is dictated by:

$$\tau_s \frac{dg_s}{dt} = -g_s \tag{6}$$

The excitatory postsynaptic potentials (EPSPs) embody an $alpha$-amino-3-hydroxy-5-methylisoxazole-4-propionic acid (AMPA)-like component alongside an N-methyl-D-aspartic acid (NMDA)-like component, taking cues from zebrafish findings [3,4]. Conversely, the inhibitory postsynaptic potentials (IPSPs) act on a single glycinergic (GLYC)-like component.

**Muscle Cells Dynamics:** Muscle cells (MC) represent an additional non-spiking population that transforms the discrete spiking trains coming from the simulated neurons into a continuous output signal which is then used to compute the output torque (see following sections). Their evolution is described by

$$\tau_{mc} \frac{dM}{dt} = -M + RI_{syn} \tag{7}$$

$$I_{syn} = \sum_{s} g_s \left(E_s - M\right) \tag{8}$$

Where $\tau_{mc}$ is the membrane time constant, $M$ is the membrane potential and the synaptic current is computed analogously to Eq. 5. Unlike the network neurons, the muscle cells do not have a spike detection and reset dynamics. Effectively, the muscle cells are modeled as firing rate neurons [1] that act as low-pass filters of the input activity. Their potential is normalized in the range $[0.0,1.0]$ and can be interpreted as the degree of activation of the motor neurons in the corresponding network region. The parameters for the muscle cells are reported in Table 1.

**Table 1. Muscle cells parameters**

| $\tau_{mc}(ms)$ | $E_{ex}$ | $E_{in}$ | $\tau_{ex}(ms)$ | $\tau_{in}(ms)$ |
|---|---|---|---|---|
| 100.0 | 1.0 | 0.0 | 2.0 | 2.0 |

**2) Network architecture**

The zebrafish locomotor network has been shown to comprise three different sub-modules, each one specialized for a different frequency range [5-8]. The three modules show a recurrent and hierarchical connectivity, producing a gear shift mechanism to activate the module that is most suited for the current locomotor speed [4]. In this work, we focused our modeling efforts to reproduce the slow module, generating frequencies below $4Hz$. Interestingly, the slow module comprises the majority of neurons within the biological network [7,8].

The simulated network comprises 5 different classes of neurons, namely excitatory V2a neurons, inhibitory V0d neurons, motoneurons (MN), reticulospinal neurons (RS) and propriosensory neurons (PS). The parameters of the different neuronal classes were derived from experimental observations in adult zebrafish (see Table 2). The neural network was simulated with the Brian2 library [9,10] using Euler integration with a timestep of 1ms.

The axial network is divided into 32 segments, according to the average number of ventral roots in adult zebrafish [11]. Each segment comprises 32 V2a neurons and 48 V0d neurons, representing the central pattern generator (CPG) component of the network [4,6,7]. Additionally, each segment includes 60 MN and 4 PS neurons [6,8,12]. The reticulospinal population comprises 100 neurons that project uniformly to the CPG network, introducing the descending drive that activates the locomotor rhythms [13].

The neuronal parameters for the different neuronal populations are listed in Table 2. The values enclosed by parentheses are sampled randomly from a uniform distribution for each neuron of the network. Notably, the properties of V2a neurons were tuned to match the spiking properties reported for the slow V2a module in [4,6]. In particular, the selected neural properties lead to 80% of the V2a neurons showing bursting properties in the $2-4Hz$ frequency range. Conversely, the remaining neurons show tonic firing. The V0d neuronal properties were tuned similarly to obtain both bursting (40%) and tonic (60%) firing patterns within the population. The RS and PS populations were modeled with non-adaptive neurons as they mostly display tonic firing properties [12,13].

**Table 2. Neuronal parameters**

| Parameter | V2a [3,4,6] | V0d [7] | RS [13] | MN [5,6] | PS [12] |
|---|---|---|---|---|---|
| $t_{\mathrm{refr}}$ (ms) | 5.0 | 2.0 | 5.0 | 5.0 | 5.0 |
| $\tau_{\mathrm{m}}$ (ms) | 10.0 | 10.0 | 26.8 | 10.0 | 14–16 |
| $R_{\mathrm{m}}$ (GΩ) | [0.1–0.9] | [0.2–0.3] | 0.16 | [0.2–0.3] | [0.14–0.16] |
| $E_{\mathrm{rest}}$ (mV) | $[-64, -56]$ | $[-64, -60]$ | -58.0 | $[-57, -55]$ | $[-58, -57]$ |
| $E_{\mathrm{thres}}$ (mV) | -20 | -20.0 | -45.0 | -20 | $[-43, -42]$ |

| Parameter | V2a [3,4,6] | V0d [7] | RS [13] | MN [5,6] | PS [12] |
|---|---|---|---|---|---|
| $E_{\mathrm{reset}}$ (mV) | -32.5 | -40.5 | -58.0 | -38 | -80.0 |
| $E_{\mathrm{rh}}$ (mV) | $[-41, -36]$ | $[-46, -40]$ | - | $[-43, -41]$ | - |
| $\Delta_T$ (mV) | 5 | 5.0 | - | 5 | - |
| $\tau_\omega$ (ms) | 500 | 700 | - | 150 | - |
| $\Delta\omega$ (pA) | 10.0 | 13.0 | 0.0 | 14.0 | 0.0 |
| $a_\omega$ (nS) | 0.1 | 0.5 | 0.0 | 2.0 | 0.0 |

Within the axial network, the connection probability between a pre-synaptic neuron with axial coordinate $x_{pre}$ and a post-synaptic neuron with axial coordinate $x_{post}$ is defined by an asymmetric exponential distribution as

$$P(x_{pre}, x_{post}) = \begin{cases} Ae^{-\frac{(x_{pre}-x_{post})^2}{4\sigma_{up}^2}}, & \text{if } x_{pre} >= x_{post} \\ Ae^{-\frac{(x_{pre}-x_{post})^2}{4\sigma_{dw}^2}}, & \text{if } x_{pre} < x_{post} \end{cases} \quad (9)$$

Where $A$ specifies the maximum connection probability, while the parameters $\sigma_{up}$ and $\sigma_{dw}$ determine how wide the probability distributions are for ascending (i.e. rostrally oriented) and descending (i.e. caudally oriented) connections respectively. The parameters for the different synaptic connections are listed in Table 3.

**Table 3. Synaptic connectivity**

| Connection | A | $\sigma_{up}$ (seg) | $\sigma_{dw}$ (seg) |
|---|---|---|---|
| V2a → V2a | 0.7 | 0.175 | 0.875 |
| V2a → V0d | 0.7 | 0.175 | 0.875 |

| Connection | A | $\sigma_{up}$ (seg) | $\sigma_{dw}$ (seg) |
|---|---|---|---|
| V0d → V2a | 0.7 | 0.35 | 1.05 |
| V0d → V0d | 0.7 | 0.35 | 1.05 |
| RS → V2a | 0.5 | - | - |
| RS → V0d | 0.5 | - | - |
| V2a → MN | 0.75 | 0.35 | 1.4 |
| V0d → MN | 0.35 | 0.7 | 1.4 |
| PS → V2a | 0.5 | 3.0 | 0.0 |
| MN → MC | 1.0 | 0.8 | 0.8 |

The V2a neurons send ipsilateral projections to the other axial neurons, while also forming highly recursive connections with other V2a units. The V0d neurons connect to contralateral V2a and V0d neurons. Additionally, the connections of V2a and V0d neurons are biased towards more caudally located neurons, as suggested by experimental observations [3,4,7]. Conversely, the axial MN populations project symmetrically to the closest neighbouring muscle cells, thus providing an estimate of the activity of the local axial region. Finally, the PS neurons target contralateral V2a neurons in the rostral direction, with projections spanning over several segments [12].

In Table 4, the parameters for the synaptic variables of the model are reported (see Eq. 5 and 6).

**Table 4. Synaptic parameters**

| Connection | $E_s$(mV) | $\tau_s$(ms) |
|---|---|---|
| Ampa | 0.0 | 20.0 |
| NMDA | 0.0 | 20.0 |
| Glyc | -65.0 | 20.0 |

The connection weights between neuronal populations were selected to obtain biologically realistic open-loop oscillations under descending drive from the RS neurons. This calibration was performed once and restricted to the isolated neural controller: no network parameter was subsequently changed or tuned for the closed-loop simulations, i.e., those including sensory feedback from the mechanical model. Table 5 reports the synaptic weights for each connection type. Notably, these values lead to post-synaptic potentials compatible with experimental observations in [4,6,12].

**Table 5. Synaptic weights**

| Connection | $\Delta g_{ampa}$ | $\Delta g_{nmda}$ | $\Delta g_{glyc}$ |
|---|---|---|---|
| $V2a \rightarrow MN$ | 0.04000 | 0.04000 | - |
| $V0d \rightarrow MN$ | - | - | 0.00500 |
| $V2a \rightarrow V2a$ | 0.00770 | 0.00770 | - |
| $V2a \rightarrow V0d$ | 0.10000 | 0.10000 | - |
| $V0d \rightarrow V2a$ | - | - | 0.00425 |
| $V0d \rightarrow V0d$ | - | - | 0.00050 |
| $RS \rightarrow V2a$ | 0.01850 | 0.01850 | - |
| $RS \rightarrow V0d$ | 0.00660 | 0.00660 | - |
| $PS \rightarrow V2a$ | - | - | 0.10000 |

**3) Mechanical model**

The mechanical model consists of a juvenile zebrafish body with a length of $1.8cm$. The axial configuration is a chain of 16 rigid links, interconnected by 15 muscle-controlled joints operating in the horizontal plane. The mechanical model was simulated in synchrony with the neural model using the MuJoCo physics engine [14] via the FARMS simulation framework [15].

**Hydrodynamic model:** When submerged in water, each link of the body is subjected to buoyancy and hydrodynamic forces. The density of the links is constant throughout the body and equal to $1000.0 kg/m^3$, rendering the model neutrally buoyant. Additionally, since adult zebrafish typically swim in high Reynolds number regimes [16], we opted for a hydrodynamic model composed of inertial drag forces. For each link, a speed dependent drag force ($F_d$) is applied in each dimension according to:

$$F_d = \frac{1}{2} \rho C_d A (V_{body} - V_{fluid})^2 \quad (10)$$

Where $A$ is the link's cross-sectional area, $c_d$ is the drag coefficient, $\rho$ is the density of water, $V_{body}$ is the link's speed and $V_{fluid}$ is the fluid speed in correspondence of the link's center of mass. The drag coefficients were calibrated to match the swimming performance reported in [17] when replaying the same kinematics in position control.

**Muscle model:** The activity of muscle cells is transduced into output torques for the mechanical model using the muscle model proposed by Ekeberg [18]. For each joint, the model receives as input the flexor ($M_L$) and extensor ($M_R$) activations from the muscle cells (see Eq. 7), as well as the current joint angle ($\theta$) and speed ($\dot{\theta}$) to compute the resulting output torque ($\tau$) via:

$$\begin{aligned} \tau = \alpha M_{diff} + \beta(\gamma + M_{sum})(\theta - \theta_0) + \delta\dot{\theta} \\ M_{diff} = G_{MC}(M_L - M_R) \\ M_{sum} = G_{MC}(M_L + M_R) \end{aligned} \quad (11)$$

Where $\alpha$ is the active gain, $\beta\gamma$ is the passive stiffness, $\beta M_{sum}$ is the active stiffness, $\delta$ is the damping coefficient, $\theta_0$ is the resting angle. Additionally, the flexor and extensor activations are scaled by a factor $G_{MC}$ that is varied across different joints to produce the sub-carangiform kinematics observed during zebrafish swimming [17,19].

The muscle parameters were calculated with the method proposed in [20] so that, to a first approximation, all joints shared a natural frequency of $5.5 Hz$, a damping ratio of 1.0 and a zero frequency gain of $2\pi/N_{joints}$. This choice ensures that all joints operate in a similar dynamic regime, while also guaranteeing that they are able to produce oscillations in the frequency ranges displayed during slow swimming ($< 4Hz$). Indeed, lower values of resonance frequency would result in overly compliant joints and in the inability to generate rapid and coordinated movements. On the other hand, higher values of resonance frequency would lead to a body that is too stiff and reduce the capability of external factors (i.e.; inertial and hydrodynamic forces) to counteract the torques generated by the muscles [21].

**Stretch feedback:** The proprioceptive feedback neurons residing in the stretched side of the body receive an input that is proportional to the angles of the mechanical joints according to:

$$I_{PS} = max(0, G_{PS} \cdot \theta) \quad (12)$$

Where $G_{PS}$ is the gain of the mechano-electrical transduction. The angle input $\theta$ provided to each sensory neuron is obtained from the linear interpolation between the two closest mechanical joints. The value of the sensory feedback gain was chosen based on the rheobase current of the sensory neurons so the neurons start to fire when the total body curvature exceeds 10 degrees.

### 4) External bending experiments

In simulations involving external bending, the simulated muscles were silenced while the target kinematics was applied in position control. In accordance to the experimental setup, the bending signal was applied to all joints except the first head joint. During the experiments, the neural network was activated to obtain locomotor rhythms. Additionally, the network still received sensory feedback from the position-controlled simulated body.

For static bending experiments, the body joints were bent simultaneously by the same angle to obtain a total tail deflection of 20 degrees lasting for 5 seconds. Similarly, in fictive pacing experiments the body joints were simultaneously bent to obtain a rhythmic tail deflection with an amplitude of 20 degrees and different frequencies, higher or lower than the ongoing network oscillations. Notably, different baseline frequencies were obtained by minimal adjustment ($\pm 25\%$) of the adaptation time constant of the CPG neurons and external drive to the RS neurons (see Eq.2). Finally, for the *in-silico* schooling experiments the schooling signal was simultaneously applied to the body joints to match the total tail deflection observed experimentally. Interestingly, in agreement with the General Theory of Synchronization [22], the strength of the entrainment could be modulated by modulating the amplitude of the applied bending, the sensory feedback gain (see Eq.12) and the difference between the natural frequency of the network and the frequency of the stimulus. However, this analysis was outside of the scope of the study and was not investigated further.

In all the mentioned experiments, the simulation started with the body kept in a straight position for 5 seconds. After the bending signal, the body returned to the straight position for 5 additional seconds before the end of the experiment.

**5) Vortex field swimming**

To simulate the fluid motion generated by a leader fish in water we used the boundary data immersion method (DBIM2, [23,24]) in two dimensions $(x, y)$ implemented in https://github.com/WaterLily-jl/WaterLily.jl. This method solves the fluid-body coupling using a second order accurate scheme where the body shape and velocities are prescribed analytically. This scheme is capable of accurately simulating turbulent flows at high Reynolds numbers (up to $10^5$, the estimated Reynolds number of adult zebrafish swimming is $\sim C \cdot 10^3$ [17].

We simulated a square tank and imposed inlet flow boundary conditions at the left/right boundaries and reflective boundary conditions at the top/bottom boundaries. The size of the tank is $7.2cm$ by $7.2cm$.

For each experiment, the inlet speed is set according to the average tail beat frequency according to the frequency-speed map in [25].

In BDMI2 the fluid-body interaction is prescribed using a signed distance function (the distance from the boundary of the body and fluid interface) and body velocities (in the x and y directions, $u_{body} = 0$ and $v_{body}$, respectively). For the fish body we considered a body length $L = 1.8cm$. The fish midline is allowed to move only on the y-axis as a function of the arclength $s = clamp(x, 0, L)/L \in [0,1]$ according to:

$$y(s,t) = d(s) \cdot \sin[2\pi(ws - f(t)t)] \quad (13)$$

In this equation the function $d(s)$ describes the displacement of the body and follows the experimental data in [19]. The time-dependent frequency function $f(t)$ was extracted from a behavioral recording (the same sequence applied in the ex-vivo schooling experiments) in order to match the cycle-to-cycle variations in the tail beat frequency. The total wave lag between the head and the tail was set to 95% of the cycle duration ($w = 1.9\pi$).

The body midline equation allowed us to derive the body velocities ($u_{body} = 0$, $v_{body} = dy/dt$) and the signed distance function $d = y(s,t) - tk(s)$, where $tk(s)$ is the body thickness, which varies along the midline, and it is given by the linear interpolation of the profile of the mechanical body of the fish.

Once the vortex fields generated with the prescribed leader fish kinematics were obtained, we studied *in-silico* vortex matching by placing the simulated (network-controlled) fish behind the wake of the leader and imposing the fluid velocities generated in BDIM2 to the drag model of the follower fish (see Eq.10). Importantly, the position of the network-controlled fish was constrained at the level of the third link, corresponding to $\approx 0.25L$, a location that exhibits minimal lateral displacement during sub-carangiform swimming [19]. The constraint allows us to systematically explore the space behind the leader without affecting the generated kinematics. The longitudinal distance between the leader and the follower was varied in the range $[0.4, 1.6]BL$. The transversal distance was varied in the range $[-0.2, +0.2]BL$. For each position, 5 different individuals (i.e.; networks generated with different seeds for the random number generation) were tested. Finally, the power consumption during *in-silico* vortex matching was evaluated by simulating 50 different individuals and testing them in open-loop and closed-loop, with and without vortices.

To test whether vortex phase matching depended on the specific choice of body stiffness, we repeated the analysis for different muscle parameter values. The control parameter was the natural frequency FN of the body muscles (see above), which was varied from 2.5 Hz (very compliant body) to 15.0 Hz (very stiff body). The damping ratio and the zero-frequency gain were kept constant across conditions, so that the muscle parameters (α, β, δ) scale with FN, with higher values corresponding to stiffer muscles. In the highly compliant condition, the muscles were overpowered by the vortices: the model could not generate forward thrust and passively followed the flow, with or without proprioceptive feedback. In the stiff condition, the model was largely insensitive to the vortices, and phase matching was strongly compromised even in closed loop. Between these two extremes there was a broad regime, including the muscle stiffness used in our study, in which vortex matching occurred only in closed loop. This shows that phase matching depends on proprioceptive feedback rather than on vortices dominating muscle torque.

**6) Metrics computation**

**Power consumption:** Power consumption ($P$) in the simulated model was estimated from joint torques $\tau_i$ and angular velocities $\dot{\theta}_i$ as

$$P = \frac{\Delta t}{T} \sum_{k=1}^{N_t} \sum_{i=1}^{N_j} \left[\tau_i(k)\dot{\theta}_i(k)\right]^+ \tag{14}$$

where $\Delta t$ is the simulation time step, $T$ is the total duration of the mechanical simulation, $N_t$ is the number of time steps, $N_j$ is the number of actuated joints, and $[x]^+ = \max(x, 0)$. Notably, we retained only the positive part of the mechanical power, corresponding to periods in which the muscles inject energy into the system. The negative part was discarded because negative mechanical work has a substantially lower metabolic cost than positive work [26,27]. Similar assumptions have been used in computational studies of swimming animals [21,26,28,29].

**Phase difference:** Analogously to the real-fish experiments, the evolution of the head-tail angle was computed by summing the joint angles along the body. The cycle phase was then calculated by mapping consecutive peaks of the trace to an interval of $[0,2\pi]$. The phase of the leader fish was computed from the experimental recording that was used to obtain the vortex field evolution (see Eq.13). Notably, the leader signal displays a time-varying frequency, and consequently, a nonlinear phase evolution. The follower phase was computed based on the evolution of the joint angles during the simulation of the network-controlled body.